\documentclass[aip,jcp,reprint,superscriptaddress,floatfix,twocolumn]{revtex4-1}
\usepackage{verbatim}
\usepackage[english]{babel}
\usepackage{amsmath, bm}
\usepackage{amssymb}
\usepackage{epstopdf}
\usepackage{comment}
\usepackage{setspace}
\usepackage{microtype}
\usepackage{rotating}
\usepackage{amstext}
\usepackage{dcolumn}

\usepackage{epsfig}
\usepackage{graphicx}
\usepackage{multirow}
\usepackage{array}
\usepackage{caption}
\usepackage{rotating}
\usepackage{keyval}
\usepackage{ifthen}
\usepackage{chemfig}
\usepackage{subfig}
\usepackage{siunitx}
\usepackage[font=small]{caption}
\usepackage[normalem]{ ulem }
\usepackage{soul}

\includecomment{reprint}
\excludecomment{preprint}

\usepackage{xcolor}
\definecolor{dgreen}{rgb}{0,.5,0}
\definecolor{dred}{rgb}{.7,.0,.0}

\usepackage{floatrow}
\usepackage[colorlinks,bookmarks=false,citecolor=blue,linkcolor=red,urlcolor=blue]{hyperref}

\begin{document}

\title{Stochastic description of the stationary Hall effect}

\author{Pierre-Michel D\'{e}jardin}
\affiliation{Laboratoire de Math\'{e}matiques et Physique, Universit\'{e} de Perpignan Via Domitia, 52 avenue Paul Alduy, F-66860 Perpignan cedex, France}

\author{Jean-Eric Wegrowe}
\affiliation{LSI, Ecole Polytechnique,CEA/DRF/IRAMIS,CNRS, Institut Polytechnique de Paris, F-91128 Palaiseau, France}

\begin{abstract}

The properties which characterize the stationary Hall effect in a Hall bar are derived from the Langevin equations describing the Brownian motion of an ensemble of interacting moving charges in a constant externally applied electromagnetic field. It is demonstrated that a non-uniform current density a) superimposes on the injected one, b) is confined in a boundary layer located near the edges over the Debye-Fermi length scale c) results from the coupling between diffusion and conduction and d) arises because of charge accumulation at the edges. The theory can easily be transposed to describe the Hall effect in metals, semi-conductors and plasmas and agrees with standard and previously published results.

\end{abstract}
\maketitle

\section{Introduction}
The physical mechanisms underlying the classical Hall effect \cite{Hall1879AmJMath} are well-known and are described in reference textbooks. For an ideal Hall bar - i.e.  a long, thin, and narrow conducting layer contacted to an electric generator with symmetric lateral edges (see Fig.1) -  the stationary state under a static magnetic field is usually defined by an accumulation of electric charges at the edges (responsible for the Hall voltage), and zero lateral current. The charge accumulation occurs over a typical scale which is the Debye-Fermi length.

However, the nature of the boundary conditions at the edges is problematic since the accumulation of electric charges is not due to an external constraint (like e.g. for a capacitor or in a simple Ohmic conductor \cite{Jackson1996AJP}). Rather, this charge accumulation is produced by the system itself in reaction to the presence of a static magnetic field (which corresponding force produces no work), in agreement with the Le Chatelier-Braun principle.\cite{Ehrenfest1911ZPhysChem}

Recently, these boundary conditions have been studied via a phenomenological basis using the least dissipation principle of non-equilibrium Thermodynamics. \cite{Creff2020JAP}. This treatment shows that a non-uniform current density flows in the direction of the injected current (see Fig.1). The surface current allows the charge accumulation to be renewed permanently, despite the absence of transfer of electric charges from one edge to the other. As a consequence, a measure of the Hall voltage with a realistic voltmeter (i.e. with a small charge leakage) does not destroy the charge accumulation. Although rather intuitive, the presence of this Hall current does not seem to have been described before.

The purpose of this work is to provide a stochastic description of the (time-independent) stationary Hall effect by formulating the problem in the general context of the modern theory of Brownian motion \cite{Coffey2017Book,Risken1989Book}, where drift and diffusion currents appear naturally as a consequence of the description of a stationary stochastic process in Fokker-Planck terms. The equations which will be employed throughout this work are similar in mathematical form to those describing diffusion of a plasma across an imposed electromagnetic field (see Ref \citenum{Jimenez2006PRE} and References therein).



\begin{figure} [h!]
   \begin{center} 
   \begin{tabular}{c}
\includegraphics[height=4cm]{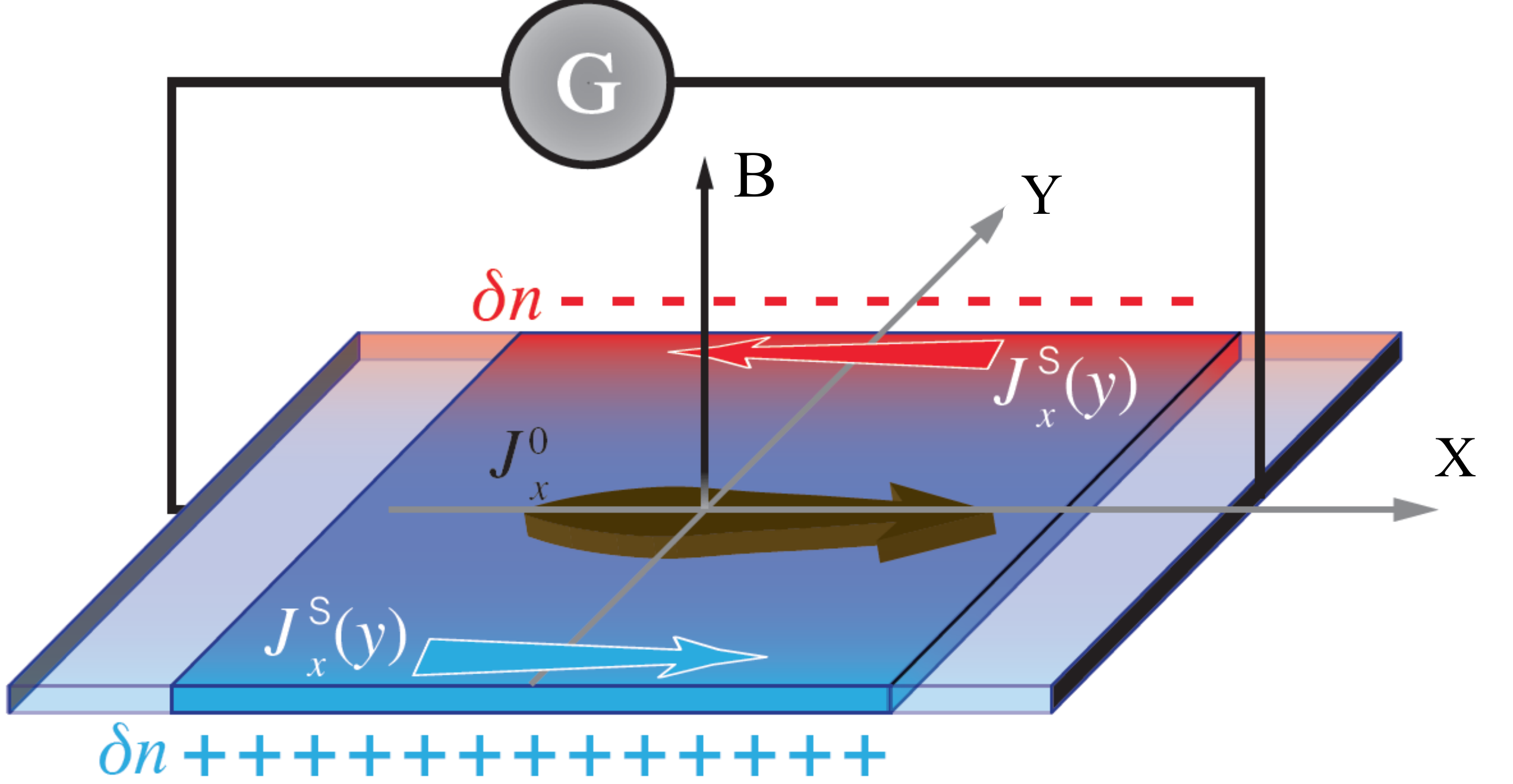}
   \end{tabular}
   \end{center}
\caption[Fig1]
{ \label{fig:Fig1} Schematic representation of the Hall effect under a static magnetic field $B$ applied along the $Z$ direction, with the electrostatic charge accumulation $\delta n$ and inhomogeneous surface currents $J_x^S(y)$ confined at the edges.}
\end{figure}

\section{Theoretical Background}

We start by considering the Langevin equations of motion for the position $\mathbf r$ and momentum $\mathbf p$ describing Brownian motion of a point charge $q$ having a mass $m$ in an electromagnetic field, viz.
\begin{eqnarray}
\frac{d\mathbf{r}}{dt}(t)=\frac{\mathbf{p}(t)}{m},
\label{Position}
\end{eqnarray}
\begin{eqnarray}
\nonumber
\frac{d\mathbf{p}}{dt}(t)&=&-q\nabla\Phi(\mathbf{r},t)+\frac{q}{m}(\mathbf{p}(t)\times\mathbf{B})\\
&-&\zeta\mathbf{p}(t)+\bm{\lambda}(t).
\label{Momentum}
\end{eqnarray}
Here, $\bf B$ is the externally applied magnetic field (supposed uniform here since we neglect the magnetic properties), $-\nabla\Phi$ is the electric part of the electromagnetic field describing both the externally applied field and charge-charge interactions, $\zeta$ is a phenomenological damping coefficient, and $\bm\lambda (t)$ is a Gaussian white noise force having the properties (see Refs. [\citenum{Coffey2017Book,Risken1989Book}] for details)
\begin{eqnarray}
\overline{\bm\lambda (t)}=\bf{0}
\label{NoiseZeroMean1}
\end{eqnarray} 
\begin{eqnarray}
\overline{\lambda_{i}(t)\lambda_{j}(t')}=2mkT\zeta\delta_{ij}\delta(t-t')
\label{FDRelation1}
\end{eqnarray}
where the overbar denotes a statistical average over the distribution of realizations of the white noise \cite{Coffey2017Book,Risken1989Book}, $\delta_{ij}$ is Kronecker's delta and $\delta(t-t')$ the Dirac delta function. Since in this work we are concerned with the steady-state situation, it is sufficient to consider the overdamped limit of Eq. \eqref{Momentum} only. Following the adiabatic elimination procedure described by Risken \cite{Risken1989Book}, in Eq. \eqref{Momentum} we take
\begin{eqnarray}
\frac{d\bf{p}}{dt}(t)=0.
\label{Adiabatic}
\end{eqnarray}
Since $\bf B$ is constant in direction and magnitude, we assume that this vector is applied along the $Z$ axis of the laboratory frame $OXYZ$. Next, we introduce the angular frequency 
\begin{eqnarray}
\omega=\frac{qB}{m}
\label{Omega}
\end{eqnarray}
where $B$ is the magnetic field modulus, and the mobility tensor
\begin{eqnarray}
\bm{\eta}_{e}=R\left(\begin{matrix}
\eta & \eta_{H} & 0 \\ 
-\eta_{H} & \eta & 0 \\ 
0 & 0 & \eta
\end{matrix} \right)
\label{MobilityTensor}
\end{eqnarray} 
where
\begin{eqnarray}
R=\frac{\eta^{2}}{\eta^{2}+\eta_{H}^{2}},
\label{R}
\end{eqnarray}
\begin{eqnarray}
\eta=(m\zeta)^{-1}
\label{OhmicMobility}
\end{eqnarray}
is the bare Ohmic mobility of charge carriers, and 
\begin{eqnarray}
\eta_{H}=qB\eta^{2}
\label{HallMobility}
\end{eqnarray}
is the Hall mobility. On using Eq.\eqref{Position} and Eqs.\eqref{Adiabatic}-\eqref{HallMobility}, Eq.\eqref{Momentum} may be rewritten as follows 
\begin{eqnarray}
\frac{d\mathbf{r}}{dt}(t)=-q\bm{\eta}_{e}\cdot\nabla\Phi(\mathbf{r},t)+\bm{\kappa}(t)
\label{LangevinEqProb}
\end{eqnarray}
where $\bm{\kappa}(t)$ is a white noise velocity defined in terms of the white noise force $\bm{\lambda}(t)$ by
\begin{eqnarray}
\bm{\kappa}(t)=\bm{\eta}_{e}\cdot\bm{\lambda}(t)
\label{NoiseVelocity}
\end{eqnarray}
the statistical properties of which can easily be deduced from Eqs.\eqref{NoiseZeroMean1} and \eqref{FDRelation1} by expanding Eq.\eqref{NoiseVelocity} in Cartesian components. Hence, using Eqs. \eqref{FDRelation1} and \eqref{NoiseVelocity} we have thee fluctuation-dissipation relations
\begin{eqnarray}
\overline{\kappa_{i}(t)\kappa_{j}(t')}=0,\quad i\neq j,
\label{FDdiffcomponents}
\end{eqnarray}
\begin{eqnarray}
\overline{\kappa_{X}(t)\kappa_{X}(t')}&=&\overline{\kappa_{Y}(t)\kappa_{Y}(t')}=2D_{e}\delta(t-t')\label{FDXYcomponents}\\
\overline{\kappa_{Z}(t)\kappa_{Z}(t')}&=&2D_{e}R\delta(t-t')\label{FDZcomponents}
\end{eqnarray}
where $D_{e}$ is an effective diffusion coefficient given by
\begin{eqnarray}
D_{e}&=&DR,\label{EffectiveDiffusionCoefficient}
\end{eqnarray}
with
\begin{eqnarray}
D&=&kT\eta\label{OrdinaryDiffusionCoefficient}
\end{eqnarray}
and $D$ is the usual diffusion coefficient of the Smoluchowski-Kramers theory \cite{Kramers1940Physica} being linked to the Ohmic mobility by Einstein's relation. Owing to all what preceeds, one may derive the Fokker-Planck equation for the density of carriers $n(\mathbf{r},t)$ corresponding to Eq.\eqref{LangevinEqProb} using the methodology described in Risken's book \cite{Risken1989Book}. This equation is
\begin{eqnarray}
\nonumber
\frac{\partial n}{\partial t}(\mathbf{r},t)=DR\nabla\cdot\left(q\beta n(\mathbf{r},t)\bm{\Theta}\cdot\nabla\Phi(\mathbf{r},t)+\mathfrak{D}\cdot\nabla  n(\mathbf{r},t)\right)\\
\label{FPEGen}
\end{eqnarray}
where $\beta^{-1}=kT$ and where we have introduced the dimensionless tensors
\begin{eqnarray}
\bm{\Theta} = \left(\begin{matrix}
1 & \theta_{H} & 0 \\ 
-\theta_{H} & 1 & 0 \\ 
0 & 0 & 1
\end{matrix} \right), \qquad\theta_{H}=\frac{\eta_{H}}{\eta},
\label{TensorTheta}
\end{eqnarray}
and
\begin{eqnarray}
\mathfrak{D}=\left(\begin{matrix}
1 & 0 & 0 \\ 
0 & 1 & 0 \\ 
0 & 0 & R
\end{matrix} \right)
\label{DiffTensor}
\end{eqnarray}
We note that Eq.\eqref{FPEGen} has the form of a continuity equation in which the drift (conduction) $\mathbf{j}_{c}(\mathbf{r},t)$ and diffusion $\mathbf{j}_{d}(\mathbf{r},t)$ currents are respectively given by
\begin{eqnarray}
\mathbf{j}_{c}(\mathbf{r},t)=-q\eta Rn(\mathbf{r},t)\bm{\Theta}\cdot\nabla\Phi(\mathbf{r},t)
\label{Conduction}
\end{eqnarray}
\begin{eqnarray}
\mathbf{j}_{d}(\mathbf{r},t)=-DR\mathfrak{D}\cdot\nabla n(\mathbf{r},t)
\label{Diffusion}
\end{eqnarray}
and the total current $\mathbf{j}(\mathbf{r},t)$ is the vector sum of Eqs.\eqref{Conduction} and \eqref{Diffusion}, so that we have
\begin{eqnarray}
\mathbf{j}(\mathbf{r},t)=\mathbf{j}_{c}(\mathbf{r},t)+\mathbf{j}_{d}(\mathbf{r},t)
\label{TotalCurrent}
\end{eqnarray}

We now \textit{assume} that $\theta_{H}^{2}<<1$. In this situation $R\approx 1$ and the tensor \eqref{DiffTensor} may be replaced by unity, while the tensor \eqref{TensorTheta} remains \textit{unchanged}. In this approximation, the Fokker-Planck equation \eqref{FPEGen} becomes
\begin{eqnarray}
\nonumber
\frac{\partial n}{\partial t}(\mathbf{r},t)&=&D\nabla\cdot\left(\nabla n(\mathbf{r},t)+q\beta n(\mathbf{r},t)\nabla\Phi(\mathbf{r},t)\right)\\
&+&\beta Dq\theta_{H}\mathbf{e}_{Z}\cdot(\nabla n(\mathbf{r},t)\times\nabla\Phi(\mathbf{r},t))
\label{FPEfinal}
\end{eqnarray}
where $\mathbf{e}_{Z}$ is a unit vector along the $Z$ axis. After this lengthy derivation we are ready to characterize the stationary state of interest.

Indeed, for the stationary state Eq.\eqref{FPEfinal} reduces to
\begin{eqnarray}
\nonumber
&\nabla &\cdot\left(\nabla n(\mathbf{r})+q\beta n(\mathbf{r})\nabla\Phi(\mathbf{r})\right)\\
&+&\beta q\theta_{H}\mathbf{e}_{Z}\cdot(\nabla n(\mathbf{r})\times\nabla\Phi(\mathbf{r}))=0.
\label{FPEStationaryGeneral}
\end{eqnarray}
which solution is
\begin{eqnarray}
n(\mathbf{r})=n_{0}e^{-q\beta\Phi(\mathbf{r})}
\label{nR}
\end{eqnarray}
where $n_0$ is the intrinsic charge carriers density, as can be checked by insertion. We note in passing that because of Eq.\eqref{nR}
\begin{eqnarray}
\nabla n(\mathbf{r})=-q\beta n(\mathbf{r})\nabla\Phi(\mathbf{r})
\end{eqnarray}
so that $n(\mathbf{r})$ as given by Eq. \eqref{nR} cancels the two terms of the left hand side of Eq.\eqref{FPEStationaryGeneral} \textit{separately}. Therefore, the simpler equation
\begin{eqnarray}
\nabla\cdot\left(\nabla n(\mathbf{r})+q\beta n(\mathbf{r})\nabla\Phi(\mathbf{r})\right)=0.
\label{FPESpecial}
\end{eqnarray}
is sufficient for describing the steady state density of charge carriers.

In order to make further progress, it is necessary to specify $\Phi(\mathbf{r})$. Here, we consider that this quantity is made of two terms. The first one arises from the voltage imposed by the experimentalist to the device. This voltage creates a uniform electric field (scalar potential gradient) $\mathbf{E}_{0}=-\nabla V_{0}(\mathbf{r})$ inside the material. The second term $V_{int}(\mathbf{r})$ arises from the charge-charge interaction and can  be written on fairly general grounds as follows :
\begin{eqnarray}
q\nabla V_{int}(\mathbf{r})=\int_{\mathfrak{V}}\nabla U_{int}(\mathbf{r},\mathbf{r}')n(\mathbf{r}')g(\mathbf{r},\mathbf{r}')d\mathbf{r}'
\label{InteractionsGeneral}
\end{eqnarray}
where $\mathfrak{V}$ is the volume of the sample, $U_{int}$ describes the interactions between pairs of relevant entities in the sample and $g$ is the pair distribution function \cite{Hansen2006Book}. However, for our purposes, this level of generality is unnecessary. In the following, we use the mean field approximation, which consists of neglecting density correlations altogether (this is similar with the Vlasov treatment of density fluctuations in a plasma \cite{Hansen2006Book}). This means that $g$ may be replaced by unity. Then, we write
\begin{eqnarray}
n(\mathbf{r})=n_{0}+\delta n(\mathbf{r})
\label{RPA}
\end{eqnarray}
where $\delta n(\mathbf{r})<<n_{0}$ describes a small departure of the charge density from uniformity. Since $\mathbf{r}$ denotes the position of a point inside $\mathfrak{V}$, altogether Eq.\eqref{InteractionsGeneral} becomes
\begin{eqnarray}
q\nabla V_{int}(\mathbf{r})\approx\int_{\mathfrak{V}}\nabla U_{int}(\mathbf{r},\mathbf{r}')\delta n(\mathbf{r}')d\mathbf{r}'
\label{InteractionsSimple}
\end{eqnarray}
Now, using
\begin{eqnarray}
U_{int}(\mathbf{r},\mathbf{r}')=\frac{q^{2}}{4\pi\varepsilon\vert\mathbf{r}-\mathbf{r}'\vert}
\label{Uint}
\end{eqnarray}
where $\varepsilon$ is the absolute static electric permittivity of the material, we have
\begin{eqnarray}
\nabla^{2}U_{int}(\mathbf{r},\mathbf{r}')=-\frac{q^{2}}{\varepsilon}\delta(\mathbf{r}-\mathbf{r}')
\label{DeltaUint}
\end{eqnarray}
where $\delta(\mathbf{r}-\mathbf{r}')$ is the three-dimensional Dirac delta function. By combining Eqs. \eqref{InteractionsSimple} and \eqref{DeltaUint}, we finally have the mean-field Poisson equation for $V_{int}$, namely
\begin{eqnarray}
\nabla^{2}V_{int}(\mathbf{r})\approx -\frac{q}{\varepsilon}\delta n({\mathbf{r}})
\label{VintFinal}
\end{eqnarray}
In the following, we shall retain only terms linear in $\delta n(\mathbf{r})$. By combining Eqs.\eqref{FPESpecial}, \eqref{RPA} and \eqref{VintFinal}, we arrive at the screening equation
\begin{eqnarray}
\nabla^{2}\delta n(\mathbf{r})-\frac{\delta n(\mathbf{r})}{\lambda_{D}^{2}}\approx 0
\label{DebyeScreening}
\end{eqnarray}
where the Debye screening length $\lambda_{D}$ is given by
\begin{eqnarray}
\lambda_{D}=\sqrt{\frac{\varepsilon kT}{q^{2}n_{0}}}
\label{DebyeLength}
\end{eqnarray}
The Debye screening equation \eqref{DebyeScreening} is commonly known to occur in equilibrium situations.  

Next, we can combine Eqs. \eqref{VintFinal}-\eqref{DebyeLength} to obtain the Laplace equation
\begin{eqnarray}
\nabla^{2}(V_{int}(\mathbf{r})+\frac{kT}{qn_{0}}\delta n(\mathbf{r}))=0
\label{MainResult1}
\end{eqnarray}
which is important in what follows. 

\section{Application to the Hall bar}

We consider now a Hall bar with length $a$, width $b$ and thickness $c$ (cf. Figure 1). Since no current is expected in the direction of the magnetic field $\mathbf{B}$, we set $j_{Z}=0$ and neglect all derivatives with respect to $z$. Furthermore, we consider that $b<<a$. As a consequence, the functions $V_{int}$ and $\delta n$ have $y$ dependence only and the Laplace Eq. \eqref{MainResult1} becomes
\begin{eqnarray}
\frac{d^{2}}{dy^{2}}(V_{int}(y)+\frac{kT}{qn_{0}}\delta n(y))=0
\label{MainResult2}
\end{eqnarray}
This equation is integrated once to yield
\begin{eqnarray}
\frac{dV_{int}}{dy}(y)=-\frac{kT}{qn_{0}}\frac{d\delta n}{dy}(y)+K
\label{InteractionField}
\end{eqnarray}
where $K$ is an integration constant to be specified later. Since the electric voltage is applied by the experimentalist in the $X$ direction, we have
\begin{eqnarray}
-\nabla\Phi(\mathbf{r})=E_{0}\mathbf{e}_{X}-\frac{dV_{int}}{dy}(y)\mathbf{e}_{Y}
\label{TotalField}
\end{eqnarray} 
where $e_{X}$ and $e_{Y}$ are unit vectors along the $X$ and $Y$ axes respectively. The conduction and diffusion current densities Eqs. \eqref{Conduction} and \eqref{Diffusion} in the mean field approximation are respectively given by
\begin{eqnarray}
\mathbf{j}_{c}(\mathbf{r})=-q\eta n_{0}\bm{\Theta}\cdot\nabla\Phi(\mathbf{r})
\label{ConductionRPA}
\end{eqnarray}
\begin{eqnarray}
\mathbf{j}_{d}(\mathbf{r})=-kT\eta\frac{d\delta n}{dy}(y). 
\label{DiffusionRPA}
\end{eqnarray}
Thus, using Eqs.\eqref{TotalField}-\eqref{DiffusionRPA} $\mathbf{j}=\mathbf{j}_{c}+\mathbf{j}_{d}$ has Cartesian components given by
\begin{eqnarray}
j_{X}(y)&=&qn_{0}(\eta E_{0}-\eta_{H}K)+kT\eta_{H}\frac{d\delta n}{dy}(y)\label{JX}\\
j_{Y}(y)&=&-qn_{0}(\eta_{H}E_{0}+\eta K)=\mathrm{const}\label{JY}\\
j_{Z}(y)&=&0\label{JZ}
\end{eqnarray}
Notice that in spite of the fact that $\mathbf{j}$ is \textit{non-uniform}, \textit{still} $\nabla\cdot\mathbf{j}=0$ and the solution of this last equation is nor $\mathbf{j}=\mathbf{0}$, nor $\mathbf{j}=\mathbf{const}$! This demonstrates the non-trivial character of the current density distribution in a Hall device, where it is the device itself \textit{which generates its own boundary conditions along the $y$ axis}.

In order to determine $K$, we use the global constraints imposed by the current injection from the generator, namely, the total electric current is zero in the $Y$ direction. This implies that we neglect significant leakage currents (e.g. due to the voltmeter, or due to a lateral circuit). Therefore, the flux of the total current density through a section containing $\mathbf{E}_{0}$ vanishes. We have
\begin{eqnarray}
I_{Y}=\int_{-a/2}^{a/2}\int_{-c/2}^{c/2}\mathbf{j}\cdot\mathbf{e}_{Y}dxdz=0
\end{eqnarray}
According to Eq.(\eqref{JY}) the current density $j_{Y}$ is a constant so that the dot product under the integral is zero. Therefore,
\begin{eqnarray}
j_{Y}=0,
\end{eqnarray}
and we have
\begin{eqnarray}
K=-\theta_{H}E_{0}
\end{eqnarray}
which is the negative of the projection of the Hall electric field. Denoting the Hall field by $E_{H}$, we deduce the well-known result
\begin{eqnarray}
\frac{E_{H}}{E_{0}}=\frac{\eta_{H}}{\eta}=qB\eta
\label{HallEquation}
\end{eqnarray}
where in the last equality Eq.\eqref{HallMobility} has been used. By replacing $K$ in Eq.\eqref{JX}, we have
\begin{eqnarray}
\nonumber
j_{X}(y)&=&qn_{0}\eta(1+\theta_{H}^{2})E_{0}+kT\eta_{H}\frac{d\delta n}{dy}(y)\\
&\approx & qn_{0}\eta E_{0}+kT\eta_{H}\frac{d\delta n}{dy}(y)
\label{JXFinal}
\end{eqnarray}  
since we have assumed during our derivation that $\theta_{H}^{2}<<1$. We note that the second term is proportional to $D_{H}=\eta_{H} kT$ which has the units of a diffusion coefficient. This term appears because of coupling between conduction and diffusion.

Note that for usual laboratory magnetic fields ($B~1$T), the second term in the right-hand side of Eq.\eqref{JXFinal} is small in comparison with the first. 

\section{Comparison with irreversible thermodynamics}

The above theory may be easily connected with the irreversible Thermodynamics of Creff et al. \cite{Creff2020JAP}. In order to recover their expression for the electrochemical potential $\mu$ of Creff et al., we rewrite the FPE \eqref{FPESpecial} as follows :
\begin{eqnarray}
\nonumber
\nabla\cdot\left( n(\mathbf{r})\left(\frac{kT}{q}\nabla\ln\frac{n(\mathbf{r})}{n_{0}} +\nabla\Phi(\mathbf{r})\right)\right)=0
\end{eqnarray}
This last equation becomes finally
\begin{eqnarray}
\nabla\cdot\left(n(\mathbf{r})\nabla\mu(\mathbf{r})\right)=0
\label{MuEq}
\end{eqnarray}
if $\mu$ is written as follows : 
\begin{eqnarray}
\nonumber
\mu(\mathbf{r})&=&\frac{kT}{q}\ln\frac{n(\mathbf{r})}{n_{0}}+\Phi(\mathbf{r})\\
&=&\frac{kT}{q}\ln\frac{n(\mathbf{r})}{n_{0}}+V_{0}(\mathbf{r})+V_{int}(\mathbf{r}),
\label{MuDef}
\end{eqnarray}
which is the starting point of the non-equilibrium Thermodynamics approach used by Creff et al. We now apply this last equation to the Hall bar in the mean field approximation. By integrating Eq.\eqref{InteractionField} once and linearizing with respect to $\delta n$ we have
\begin{eqnarray}
V_{int}(y)=-\frac{kT}{qn_{0}}\delta n(y)-E_{H}y
\label{VVint}
\end{eqnarray}
where an unsignificant integration constant has been ignored. Because, in the mean field approximation, we have Eq.\eqref{RPA}, Eq.\eqref{MuDef} becomes after trivial algebra
\begin{eqnarray}
\mu(x,y)&=&-E_{0}x-E_{H}y
\label{MuResRPA}
\end{eqnarray}
since $V_{0}(\mathbf{r})=V_{0}(x)=-E_{0}x$. This definitely proves that the stationary state in the Hall bar leads to a linear variation of the electrochemical potential $\mu$ with $y$ (Creff et al. use $\tilde \mu=\mu-V_{0}$ instead in their work).

\section{Discussion}

In order to quantitatively illustrate our calculation, we consider a situation where charges are accumulated at the edges in a symmetric manner. In this situation, we have 
\begin{eqnarray}
\delta n(y)=\delta n_{a}\frac{\sinh (y/\lambda_{D})}{\sinh (b/(2\lambda_{D}))},
\label{Deltansymmetric}
\end{eqnarray} 
where $\delta n_{a}=\delta n(b/2)$ is the charge accumulated at the $y=b/2$ boundary. The symmetry of the two edges imposes $\delta n(0)=0$ and the mean number of charge carriers (corresponding to the electroneutrality) is $n_0$.

This gives for $j_{X}(y)$ the expression
\begin{eqnarray}
j_{X}(y)&=&qn_{0}\eta E_{0}+\frac{kT\eta_{H}\delta n_{a}}{\lambda_{D}}\frac{\cosh (y/\lambda_{D})}{\sinh (b/(2\lambda_{D}))}
\label{JXpositivey}
\end{eqnarray}
We are interested only in the nonuniform part of the current, viz.
\begin{eqnarray}
J_{X}^{S}(y)=j_{X}(y)-J_{0}
\end{eqnarray}
where
\begin{eqnarray}
\nonumber
J_{0}=qn_{0}\eta E_{0}
\end{eqnarray}
is the usual Ohmic current. 
in agreement with the predictions of Creff et al.\cite{Creff2020JAP} derived in the context of non-equilibrium Thermodynamics. Therefore, the present approach validates the presence of an interface current in a boundary layer of thickness $\lambda_{D}$ and more generally, the approach of nonequilibrium Thermodynamics to this problem. The situation described above is shown in Figure 2 below.


\begin{figure}[h!]
\includegraphics[width=\textwidth]{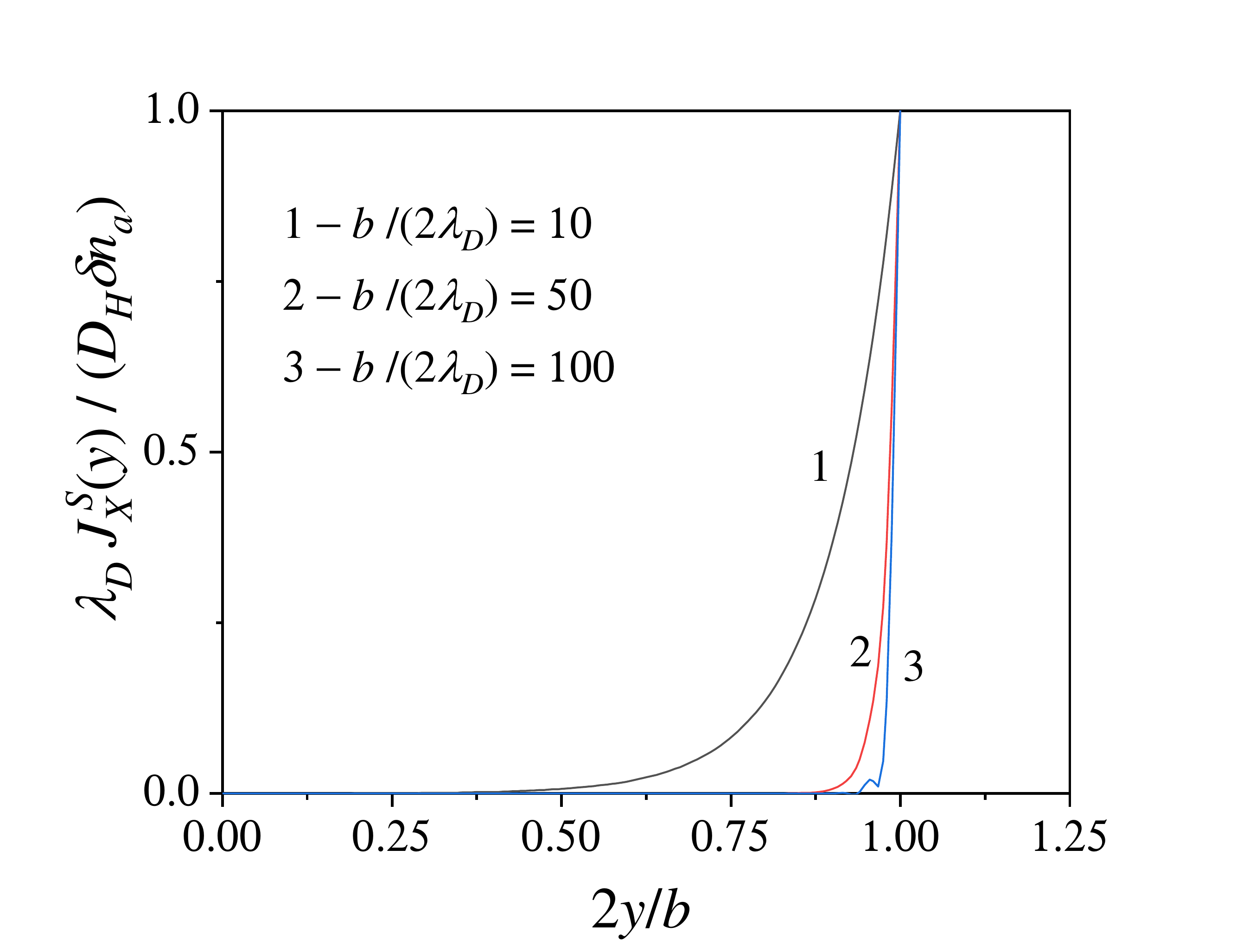}
\caption{Nonuniform part of the current density $J_{X}^{S}(y)=j_{X}(y)-J_0$ for various values $b/(2\lambda_D)$.} 
\label{fig:Fig2}
\end{figure}

The application of our theory to the Hall effect in a fully degenerated metal is obtained by substituting $T_F$ the Fermi temperature to the temperature $T$.
Since the starting equations which we use are relevant to plasma physics \cite{Jimenez2006PRE}, the present calculations can be adapted to handle the Hall effect in plasmas by simply relaxing the approximation $\theta_{H}^{2}<<1$ as the charge carriers are noticeably deviated in ionized gases. In this context, a dynamical solution might be of interest for qualitative comparison with the data of Glattli et al. \cite{Glattli1985PRL}, rather than the stationary one that we have provided here. The mathematical treatment is however much more involved.

At last, our treatment can be extended to the determination of the steady-state spin Hall effect with finite spin-flip length. In particular, it will be interesting to compare the results obtained from an adaptation of the present method with those obtained by us recently \cite{Wegrowe2018EPL}. 
\newline
   
\section{Conclusion}

In conclusion, we have proposed a theoretical treatment for handling the steady-state Hall effect by formulating the problem in terms of Brownian motion in a field of force. We have illustrated our method by determining the properties of the stationary Hall effect in a Hall bar, and revealed a nontrivial nonuniform contribution to the current density which results from the coupling between conduction and diffusion. For realistic values of $b$ and $\lambda_{D}$, this current density is confined in a boundary layer of thickness $\lambda_{D}$ near the boundaries parallel to the direction of the Ohmic current density. This model noticeably strengthens previous results \cite{Creff2020JAP} obtained from a description in terms of non-equilibrium Thermodynamics. 

\bibliographystyle{unsrt}
\bibliography{biblio}

\end{document}